# МОДУЛЬНАЯ ТЕХНОЛОГИЯ РАЗРАБОТКИ ПРОБЛЕМНО-ОРИЕНТИРОВАННЫХ РАСШИРЕНИЙ САПР РЕКОНСТРУКЦИИ ПРЕДПРИЯТИЯ

Мигунов В.В.
ЦЭСИ РТ, Казань
vmigunov@csp.kazan.ru

Рассматривается унифицированная модульная технология создания проблемно-ориентированных расширений САПР, реализованная в системе TechnoCAD GlassX для проектирования реконструкции предприятий. Модульность технологии заключается в хранении всех параметров проекта в одном элементе чертежа - модуле, с автоматической генерацией геометрической части модуля по этим параметрам. Излагаются общие принципы системной организации разработки расширений: выделение автоматизируемой части проекта, организация параметров в виде списков объектов с их свойствами, связями принадлежности, выделение общих и специальных операций, стадийность разработки, границы применимости технологии.

Эффективность САПР решающим образом зависит от адекватного представления объекта проектирования. Проекты реконструкции предприятий охватывают много небольших объектов с различных сторон - технологической, монтажной, строительной, электротехнической, санитарно-технической и др. согласно требованиям системы проектной документации для строительства (СПДС). В этих условиях единого представления объектов просто нет, и для целей автоматизации приходится набирать несколько различных САПР [1] с естественными сложностями в освоении их интерфейса и неудобством работы в разных средах, вплоть до сомнений в экономической целесообразности применения компьютеров [2] при большом числе индивидуальных объектов проектирования. Известно, что даже в машиностроении, где проектирование естественно связывается с технологической подготовкой производства и автоматизированным изготовлением деталей, возникают специализированные программы, ориентированные на решение узкого круга задач или определенную предметную область [3].

Наиболее распространена практика создания проблемно-ориентированных решений на основе AutoCAD. По сведениям [4], "по всему миру их создают более 6000 зарегистрированных фирм-разработчиков, а количество малых разработок и утилит просто не поддается учету". Естественно, не приходится говорить о единстве подходов, терминологии и пользовательского интерфейса. Использование этих наработок сталкивается с известными трудностями: несоответствие российским стандартам, необходимость работать в разных средах, англоязычные документация, справки, запросы и сообщения. Процессы адаптации к российским реалиям идут, но медленно, и в первую очередь в тех сферах, где применение САПР уже носит массовый характер. К этим сферам деятельность проектно-конструкторских подразделений (ПКО) предприятий не относится, а именно им чаще всего приходится создавать проекты реконструкции предприятий.

В настоящей работе описываются общие принципы и системные модельные представления, лежащие в основе технологии создания проблемно-ориентированных расширений САПР, разработанной в период с 1994 по 2003 г.г. в условиях наличия собственного графического ядра эволюционным путем, по мере охвата различных частей проекта. В отечественной САПР TechnoCAD GlassX, ориентированной на задачи проектирования реконструкции предприятия в собственном ПКО, максимальное единство модельных представлений и среды проектирования для различных частей проекта обеспечивается путем применения модульной технологии создания проблемно-ориентированных расширений. При создании каждого нового расширения используются уже имеющиеся наработки, обеспечивающие соответствие стандартам ЕСКД и СПДС, единый русскоязычный пользовательский интерфейс и терминологию, к которым уже привык пользователь. В основе модульной технологии создания расширений лежит понятие модуля как элемента чертежа.

Далее будут рассмотрены понятия модуля в чертеже, особенности параметрического представления объекта проектирования в модулях, классификация операций, выполняемых в расширении САПР, стадии разработки и область применимости модульной технологии.

*Модуль в чертеже*

Все специализированные параметры данной части проекта хранятся в одном элементе чертежа, который получил название модуль. В [5] охарактеризована сущность модулей в чертеже, в [6] освещены связанные с модулями возможности снижения требуемой памяти компьютера, в [7] описано использование модулей типа "Табличный" в задачах специфицирования чертежей. Модуль включает также видимую геометрическую часть, которая перегенерируется при всяком изменении параметров - параметрическая часть первична и полностью определяет геометрическую. Как элемент чертежа, он может выбираться, сдвигаться, удаляться и т.д. Структурирование параметров в модуле задается признаком его типа. Каждому типу модулей соответствует допустимое множество свойств. Имеющиеся на лето 2003 года в TechnoCAD GlassX типы модулей приводятся ниже, у части из них указаны также допустимые свойства в квадратных скобках:

- "Пользовательский" ["Привязка", "Симметрия", "Комментарий"];
- "Трубопровод";
- "Арматура" ["Привязка", "Симметрия", "Комментарий", "Строительная длина", "Обозначение", "Наименование", "Масса", "Примечание", "Dy", "Py"];
- "Таблица КИПиА";
- "Прибор" ["Привязка", "Несущая геометрия", "Позиционное обозначение", "Обозначение", "Наименование", "Масса", "Примечание", "Тип, марка оборудования", "Единица измерения", "Код единиц измерения", "Код завода-изготовителя", "Код оборудования, материала", "Цена", "Наименование и технич. х-ка", "На щите", "Функциональный признак прибора", "Верхний индекс", "Нижний индекс", "Комментарий", "Тип линии приборов КИП"];
- "Исполнительный механизм";
- "План этажа" ["Комментарий", "Параметры этажа в плане", "Масштаб при создании"];
- "Обозначение для аксонометрии" ["Привязка", "Симметрия", "Комментарий", "Вырезаемый на трубе отрезок"];
- "Аксонометрическая схема" ["Комментарий", "Параметры аксонометрич. схемы", "Масштаб при создании"];
- "Оформление чертежа" ["Описание оформления чертежа"];
- "Табличный" ["Комментарий", "Описание таблицы"];
- "Позиционное обозначение" ["Тип позиционного обозначения", "Тип объекта позиционного обозначения", "Специфицирующие свойства"];
- "Профиль наружной сети ВК" ["Комментарий", "Параметры профиля наружной сети ВК", "Масштаб при создании"];
- "Молниезащита зданий и сооружений" ["Комментарий", "Параметры

молниезащиты зданий и сооружений", "Масштаб при создании"];
- "Электронная подпись" ["Сотрудник", "Должность", "Пароль", "Дата", "Время"].

Не все перечисленные типы модулей включаются в расширения САПР, разработанные по модульной технологии. "Пользовательский", например, носит общий характер и может создаваться пользователем. Также он служит для приема атрибутов блоков при импорте чертежей из DXF формата.

*Параметрическое представление объекта проектирования в модулях*

Параметрическое представление (ПП) включает списки объектов и общие параметры, часть из которых носят смысл установок.

Списки объектов - массивы в памяти, за каждым элементом которых закрепляются все сведения об объекте. Сведения об одном объекте в зависимости от его сути включают геометрические характеристики, признаки ориентации, специфицирующие свойства, цвет, тип линии, тексты и др., включая ссылки на объекты других списков по номерам в этих списках (и только по номерам). Например, как правило, тексты в чертежах относятся к каким-то другим элементам, и возникают списки текстов со ссылками на эти объекты. Совокупность списков ПП по существу является реляционной базой данных, и должна удовлетворять условиям ссылочной целостности - не должно быть ссылок на отсутствующие элементы [8]. Номера в списках играют роль первичных ключей. Также, поскольку объекты в списках добавляются, удаляются и заменяются, к ссылкам предъявляются некоторые требования нормализации отношений [8] - они должны быть независимыми. Чаще всего для обнаружения зависимых ссылок бывает достаточно проверить отсутствие цикличности ссылок в совокупности списков. Может потребоваться и создание чисто ссылочных списков, в которых объектами являются только отношения (связи).

Кроме связей принадлежности, встречаются и более сложные связи. Например, в аксонометрических схемах для сокращения их размеров либо для наглядности допускаются разрывы труб со смещением полупространств по нормали к плоскости разрыва. Если все трубы, пересекающие плоскость разрыва, идут по нормали к ней, такое смещение возможно; в противном случае нельзя создавать новый объект с такой плоскостью разрыва в списке смещений.

Общие параметры относятся ко всему ПП. Это, например, точка привязки в чертеже, установка вычерчивания размеров по умолчанию, установка вычерчивания всех отметок высоты. Различаются установки объектов, действующие по умолчанию, и установки для всех объектов какого-либо списка. Первые действуют при добавлении нового объекта, помещаются в его свойства и в дальнейшем могут быть изменены для одного этого объекта. Вторые действуют сразу на все объекты списка и в нем уже не хранятся соответствующие свойства.

С точки зрения пользователя операции, которые ему предоставляются, могут быть весьма специальными для каждого из расширений. Но он видит их всегда сгруппированными в отдельное меню - основное меню данного расширения САПР, и быстро привыкает к их названиям и особенностям. В каждом расширении идет работа не с чертежом, а с ПП, в своем режиме. На экране постоянно поддерживается изображение, соответствующее текущему ПП, но входящие в него геометрические элементы еще не являются частью чертежа, они помещаются в чертеж временно. Это так называемые рабочие модули (типа "Пользовательский"), каждый из которых обычно соответствует одному видимому объекту. В параметрах рабочих модулей есть сведения, идентифицирующие список, номер объекта в нем, а также некоторые дополнительные сведения. Параметрическое представление наращивается и корректируется, после чего помещается в чертеж как один модуль.

*Типовые и специальные операции*

С точки зрения технологии разработки операции, выполняемые над ПП, разделяются на типовые и специальные. Типовые - те, которые встречаются в разных расширениях САПР. Их типичность определяется общностью структур ПП и позволяет говорить о наличии технологии разработки расширений САПР. К типовым операциям относятся:

1. Выбор в основном меню данного расширения САПР. В меню сразу включаются типовые опции добавления, удаления, редактирования, чтения ПП с диска, взятия ПП из модуля в чертеже, записи ПП на диск, помещения результатов в чертеж.
2. Преобразование параметрического представления из компактного формата в развернутый рабочий и обратно. В модулях чертежа, при временном резервном копировании в оперативной памяти и при хранении ПП на диске используется компактный формат с упаковкой по битам с выравниванием на границу байта [6]. Рабочее представление развернуто в памяти для быстрого доступа к его элементам и имеет дополнительные данные в списках, которые не хранятся в модуле, а служат целям ускорения различных операций (скоростные переменные).
3. Освобождение памяти, занятой рабочим представлением.
4. Запись на диск и чтение с диска комплекта параметров (ПП). Хранимые на диске комплекты параметров образуют наращиваемую информационную базу проектирования, они используются как прототипы.
5. Инициализация установок. Часть установок берутся из действующих в чертеже, часть назначаются специально для этого расширения.
6. Подсчет значений скоростных переменных. Это, например, габариты текстов, другие характеристики, которые требуют трудоемких вычислений, замедляющих перегенерацию изображений объектов при движении курсора.
7. Добавление объекта в список. Создается новая копия массива в памяти с пустым последним элементом.
8. Пересчет координат между всеми системами координат, используемыми в чертеже и в проектируемой части. В чертеже есть системы координат Натура и Бумага, а в проектируемой части иногда таких систем координат бывает много. Например, в проектах молниезащиты имеется один обязательный вид сверху и произвольное число вертикальных сечений зон защиты. Для каждого из этих видов и сечений имеется своя система координат с различными масштабами.
9. Генерация рабочих модулей для видимых объектов ПП. Эти операции реализуются с максимальными требованиями к быстродействию, так как они отрабатывают на всякое шевеление курсора пользователем при модификации ПП.
10. Перегенерация всех или подмножества рабочих модулей с заменой их в чертеже. Осуществляется обычно после каждой операции модификации ПП.
11. Выбор в чертеже одного или нескольких рабочих модулей из заданного множества списков для проведения каких-либо операций над соответствующей частью ПП.
12. Удаление объекта или множества объектов из списка. При удалении объекта автоматически удаляются ссылающиеся на него объекты других списков, производится перенумерация остающихся ссылок. При этом обеспечивается ссылочная целостность.
13. Взятие ПП из модуля в чертеже.
14. Помещение результатов в чертеж со сбором геометрических частей рабочих модулей в один, с запросом положения в чертеже, с заменой модуля в чертеже или с его добавлением.
15. Подсчет общих габаритов части проекта. Это необходимо для показа всего изображения в заданном окне, в том числе по нажатию "горячей клавиши".
16. Показ сгенерированных рабочих модулей в окне подсветки при выборе комплекта параметров для чтения с диска.

О специальных операциях в рамках общей характеристики технологии достаточно сказать, что они редко бывают похожими друг на друга, и в совокупности с параметрическим представлением задают "индивидуальность" каждого из расширений САПР. Отдельные расширения в рамках данной работы не рассматриваются.

*Технологические стадии*

Этап 1. Выделение части проекта, подготовка которой будет автоматизироваться в создаваемом расширении САПР. С одной стороны, желательно охватить как можно более широкий круг задач. С другой стороны, эти задачи должны допускать общее внутреннее представление, то есть быть близкими по сути. В некоторых случаях индивидуальность задачи настолько сильна, что выбор части проекта становится тривиальным. Это, например, проекты молниезащиты зданий и сооружений, где главная задача - автоматические расчет и построение горизонтальных и вертикальных сечений зон защиты молниеприемников [9]. В то же время аксонометрические схемы трубопроводных систем оказываются очень близкими в различных частях проекта реконструкции: это и чертежи специальных технологических трубопроводов [10], и схемы систем водопровода и канализации [11], и схемы систем отопления, теплоснабжения, вентиляции, кондиционирования воздуха [12]. На этом этапе требуется подробное изучение нормативной документации и сильное взаимодействие с потенциальными пользователями.

Этап 2. Определение состава объектов в параметрическом представлении, их свойств и их связей. Здесь производится основная работа по моделированию объекта разработки. Параметрическое представление должно отвечать перечисленным выше требованиям к ссылкам в списках и одновременно обеспечивать возможность корректного выполнения всех типовых и специальных операций.

Этап 3. Определение состава общих параметров, установок по умолчанию и общих установок. В зависимости от распределения установок в свойствах объектов в списках появляются или исчезают те свойства, которые задаются в установках.

Этап 4. В программном коде САПР создаются структуры, реализующие параметрическое представление, и программируются операции с номерами 1-8 в списке типовых операций. Появляется возможность добавлять объекты в списки (пока программным путем), сохранять ПП на диске и соответственно иметь несколько вариантов ПП. С этого момента разработка расширения уже легко распараллеливается между программистами. Следующие этапы по каждому из списков объектов выполняются в значительной степени параллельно, с согласованием по мере начала обработки новых ссылок (связей объектов). Одновременно идет доработка и уточнение ПП. Как только новые операции получают программную реализацию, они подключаются к основному меню (становятся доступными). Различные специальные операции программируются по мере возникновения необходимости.

Этап 5. Программируются интерактивные операции добавления объектов в списки. Для объектов различной природы они различны, это специальные операции.

Часто одна (с точки зрения пользователя) операция добавления приводит к добавлению нескольких объектов в несколько списков. Например, в аксонометрических схемах при добавлении пространственной ломаной - оси трубопровода появляются несколько новых точек в списке точек и несколько новых труб, каждая из которых задается ссылками на точку начала и точку конца.

Этап 6. Программируется генерация рабочих модулей для объектов в списках (операция 9). Теперь уже можно видеть добавляемые в ПП объекты и соответственно быстрее выявлять ошибки. По мере добавления операций генерации рабочих модулей для каждого из списков эти возможности добавляются в процедуру полной перегенерации (операция 10).

Этап 7. Программируется типовая функция выбора пользователем одного рабочего модуля и их множества (операция 11). От одного специализированного расширения к другому эта функция меняется незначительно, в зависимости от потребностей выбора. Выбор нужен в специальных операциях редактирования, удаления, связывания, переноса, копирования и др. Уже можно реализовывать специальные операции, опирающиеся на выбор объектов.

Этап 8. Программируется единая безынтерфейсная процедура удаления объекта заданного типа из ПП (операция 12), а затем и интерактивные операции удаления объектов или множеств объектов.

Этап 9. Программируются операции 13-16.

Этап 10. Комплексная отладка, написание текстов сообщений, запросов, контекстной справки, документации.

*Область применения технологии*

В полном систематическом виде модульная технология уже использована для создания трех проблемно-ориентированных расширений TechnoCAD GlassX, автоматизирующих подготовку:

- аксонометрических схем трубопроводных систем [13];
- чертежей профилей наружных сетей водоснабжения и канализации [14];
- проектов молниезащиты зданий и сооружений [15].

Отдельные элементы технологии по мере ее развития выявлялись в ходе применения модулей в качестве:

- средства автоматизации оформления чертежа рамками, основной и дополнительными надписями с их заполнением;
- средства хранения сведений о трубопроводе, обеспечивающего генерацию его чертежа по ломаной осевой линии и установкам (диаметр и др.) с последующей автоматической привязкой к нему модулей трубопроводной арматуры, хранящих в себе сведения о возможности точечной, осевой, угловой и тройниковой привязки [16];
- условных графических обозначений приборов и исполнительных механизмов в схемах автоматизации, чье геометрическое изображение перегенерируется всякий раз при изменении параметров [17];
- хранилища всех параметров чертежей строительной подосновы в плане и в разрезе [18, 19];
- универсальной модели табличных конструкторских документов [7, 20, 21].

Исходя из изложенных подходов модульной технологии и из опыта ее развития и использования, можно очертить следующие границы ее применимости в сфере автоматизации задач проектирования реконструкции предприятия:

- включаемые в одно расширение части проекта обладают внутренними связями, существенно более сильными, нежели их связи с остальными

частями проекта. Это собственно, принцип системного подхода, в соответствии с которым системы отделяются от окружения. Именно автоматический учет этих связей и дает эффект от создания проблемно-ориентированного расширения САПР. Вышеперечисленные применения модульной технологии дают примеры вариантов таких внутренних связей;
- расположение всего чертежа, относящегося к одному расширению (модуля), на одном листе. Практика показывает, что независимо от используемой САПР проектирование реконструкции ведется путем поочередной работы с различными листами проекта, хранимыми в различных файлах, и организация одновременной работы с ними в режиме модульной технологии затруднительна как в разработке, так и, главное, в использовании. В TechnoCAD GlassX, в частности, при специфицировании сразу нескольких листов в одном табличном модуле сами эти листы понимаются как внешние источники данных, в которые нельзя вносить изменения. Сам табличный модуль разрабатывается только в текущем чертеже;
- как можно более полный доступ разработчика к графическому ядру САПР.

Интересно, что с точки зрения возможности параметрического представления части проекта явных ограничений не видно. Вероятно, дело в том, что структура ПП вместе с типовыми и специальными операциями для конкретной САПР по функциям уже включает в себя те возможности, которые в последние годы предлагают крупнейшие производители СУБД: реляционная база данных с возможностью связи с изображениями, хранимые процедуры (или триггеры). При этом модульная технология не претендует на универсальность, защищенность при многопользовательском доступе, как СУБД, и допускает произвольный доступ к спискам объектов, а не изолирует разработчика от БД стандартами языка (SQL, ODBC и др.) [22].

*Использованные источники*

1. Орельяна И.О. Автоматизация при реконструкции и развитии промышленных объектов в России//CADmaster, 2001, № 3. – М.: Consistent Software.

2. Никуленкова Т. Т., Лавриненко Ю. И., Ястина Г. М. Проектирование предприятий общественного питания. – М.: Колос, 2000. – 216 с.

3. Селиванов С.Г., Иванова М.В. Теоретические основы реконструкции машиностроительного производства. – Уфа: Гилем, 2001 .– 312 с.

4. Орельяна И.О. САПР //CADmaster, 2001, № 5. – М.: Consistent Software.

5. Мигунов В.В. Модуль в чертеже как основа технологии разработки проблемно-ориентированных расширений САПР//Тез. докл. XII Международной конференции по вычислительной механике и современным прикладным программным системам, 30 июня - 5 июля.2003 г., Владимир, т.2. – Владимир: ВГУ, 2003, С.473-474.

6. Мигунов В.В. Модель чертежа и методы разработки САПР в условиях дефицита оперативной памяти//Тез. докл. XII Международной конф. по вычислительной механике и современным прикладным программным системам, 30 июня - 5 июля.2003 г., Владимир, т.2. – Владимир: ВГУ, 2003, С.471-472.

7. Мигунов В.В. Модель табличных конструкторских документов для работы с электронными каталогами и специфицирования в САПР//Информатизация процессов формирования открытых систем на основе САПР, АСНИ, СУБД и систем искусственного интеллекта: Материалы 2-й международной научно-технической конференции, 25-27 июня 2003 г., Вологда. – Вологда: ВоГТУ, 2003, С.153-156.

8. Дейт, К.Дж. Введение в системы баз данных.: Пер с англ. – 6-е изд. – Киев.: Диалектика, 1998. – 784с.

9. РД 34.21.122-87. Инструкция по устройству молниезащиты зданий и

сооружений. – М.: Мин-во энергетики и электрификации СССР, 1987.

10. Система проектной документации для строительства. ГОСТ 21.401-88. Технология производства. Основные требования к рабочим чертежам. – М.: Госстандарт, 1988.

11. Система проектной документации для строительства. ГОСТ 21.601-79 (1983). Водопровод и канализация. Рабочие чертежи. – М.: Госстандарт, 1983.

12. Система проектной документации для строительства. ГОСТ 21.602-79 (1981). Отопление, вентиляция и кондиционирование воздуха. Рабочие чертежи. – М.: Госстандарт, 1981.

13. Мигунов В.В., Сафин И.Т., Кафиятуллов Р.Р. Программная система подготовки аксонометрических схем трубопроводных систем TechnoCAD GlassX 10 (аксонометрия)//Компьютерные учебные программы и инновации, 2002, № 4(8). – М.: Минобраз РФ, С. 41.

14. Мигунов В.В., Кафиятуллов Р.Р., Сафин И.Т. Программная система TechnoCAD GlassX 19 (профиль) подготовки чертежей профилей наружных сетей водоснабжения и канализации//Компьютерные учебные программы и инновации, 2003, № 1. – М.: Минобраз РФ, С. 50.

15. Мигунов В.В., Кафиятуллов Р.Р., Сафин И.Т. Программная система автоматизации проектирования молниезащиты зданий и сооружений TechnoCAD GlassX версия 20 (молниезащита)//Компьютерные учебные программы и инновации, 2003, № 2. – М.: Минобраз РФ, С. 50-51.

16. Мигунов В.В., Сафин И.Т., Кудрявцев Д.А., Потапов А.Ю., Макаров А.М. Программная система автоматизированной подготовки чертежей расположения оборудования и трубопроводов TechnoCAD GlassX 07//Компьютерные учебные программы и инновации, 2002, № 2(6). – М.: Минобраз РФ, С.

17. Мигунов В.В., Потапов А.Ю., Кудрявцев Д.А., Свилин О.Ю., Сафин И.Т. Программная система автоматизированной подготовки схем автоматизации технологических процессов TechnoCAD GlassX 06 (схемы автоматизации)//Компьютерные учебные программы и инновации, 2002, № 2(6). – М.: Минобраз РФ, С.

18. Мигунов В.В., Потапов А.Ю., Детков И.С., Кудрявцев Д.А., Свилин О.Ю. Программная система генерации чертежей в плане строительной подосновы TechnoCAD GlassX 02 (подоснова в плане)//Компьютерные учебные программы и инновации, 2002, № 2(6). – М.: Минобраз РФ, С.

19. Мигунов В.В., Потапов А.Ю., Детков И.С., Кудрявцев Д.А., Свилин О.Ю. Программная система генерации чертежей в разрезе строительной подосновы TechnoCAD GlassX 04 (подоснова в разрезе)//Компьютерные учебные программы и инновации, 2002, № 2(6). – М.: Минобраз РФ, С.

20. Мигунов В.В., Сафин И.Т., Кафиятуллов Р.Р. Программная система подготовки спецификаций к аксонометрическим схемам участков трубопроводов монтажно-технологической части проектов по промышленным каталогам на давления до 100 кгс/см2 TechnoCAD GlassX 14 (спецификация МТ)//Компьютерные учебные программы и инновации, 2002, № 4(8). – М.: Минобраз РФ, С. 43.

21. Мигунов В.В., Сафин И.Т., Кафиятуллов Р.Р. Программная система подготовки спецификаций к аксонометрическим схемам участков трубопроводов монтажно-технологической части проектов по промышленным каталогам на давления свыше 100 кгс/см2 TechnoCAD GlassX 16 (спецификация ТВД)//Компьютерные учебные программы и инновации, 2002, № 4(8). – М.: Минобраз РФ, С. 44-45.

22. Коннолли Т., Бегг К., Стачан А. Базы данных: проектирование, реализация и сопровождение.: Пер с англ. – М.: Издательский дом "Вильямс", 2000. – 1120с.